\documentclass{article}
\usepackage[a4paper]{anysize}
\usepackage{amsthm,amsmath}
\usepackage{tikz,pgf}

\iffalse
\usepackage{refcheck}
\usepackage[draft]{changes}
\else
\usepackage[final]{changes}
\fi

\theoremstyle{plain}
\newtheorem{theorem}{Theorem}[section]

\title{Dynamical \replaced{boundary value problem}{contact problems} for a viscoelastic half-space with
  cut}

\date{}

\author{Shavlakadze, Nugzar\footnote{\added{Iv. Javakhishvili} Tbilisi
  State University A. Razmadze Mathematical Institute,
  email:nusha1961@yahoo.com} \and
  Odishelidze, Nana\footnote{\replaced{Iv. Javakhishvili Tbilisi
    State}{Georgian Technical} University, email:
  nana.odishelidze@tsu.ge}, \and
  Pachulia, Bachuki  \footnote{Georgian Technical University, e-mail: pachulia.b@gtu.ge}, \and
  Criado-Aldeanueva, F.\footnote{Department of Applied Physics II,
  Polytechnic School, Malaga University, email: fcaldeanueva@ctima.uma.es}}

\begin{document}
\maketitle

\begin{abstract}
  The dynamical \replaced{boundary value problem}{contact problems}
  for viscoelastic half-space with cut in the form of a strip
  \replaced{is}{are} considered. The \replaced{problem is}{ solutions
    of the problems are} reduced to the \replaced{singular integral
    equation of first kind}{integro-differential equations}. Using the
  method of orthogonal polynomials the \replaced{integral equation
    is}{integro-differential equations are} reduced to an infinite
  system of linear algebraic equations. The quasi-completely
  regularity of the obtained system is proved and the reduction method
  for approximate solution is developed.
\end{abstract}

\section{Statement of the problem}

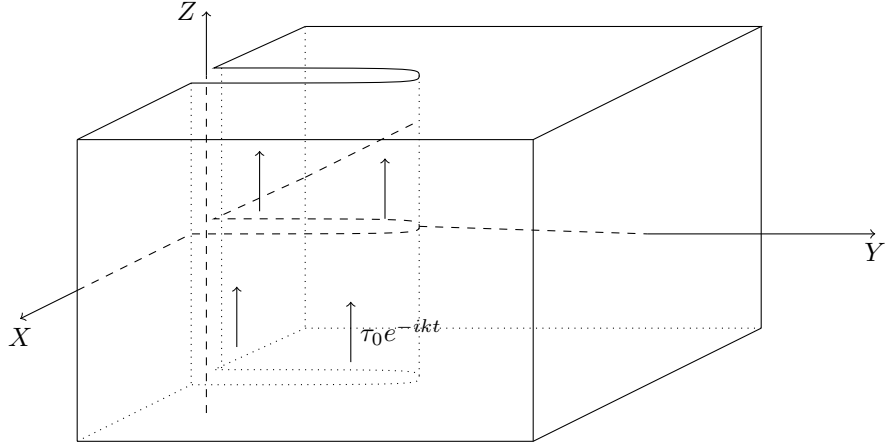
\begin{figure}
  \centering
  \newlength{\ins}
  \setlength{\ins}{3cm}
  \newlength{\insw}
  \setlength{\insw}{.4cm}
  \newlength{\wid}
  \setlength{\wid}{1.5cm}
  \newlength{\lar}
  \setlength{\lar}{6cm}
  \newlength{\hegh}
  \setlength{\hegh}{4cm}
  \begin{tikzpicture}[yscale=.5]
    \path[save path=\insertion] (-\wid,-\wid) -- (0,0) .. controls (\ins,0) .. (\ins,\insw*.5) .. controls (\ins,\insw) .. (.3,\insw) -- (\wid,\wid);
    \draw[use path=\insertion];
    \draw (\wid,\wid) -- (\lar+\wid,\wid) -- (\lar-\wid,-\wid) -- (-\wid,-\wid);

    \draw[dashed,transform canvas={yshift=-\hegh*.5},use path=\insertion];
    \draw[dashed,transform canvas={yshift=-\hegh*.5}] (\wid,\wid) -- (2*\wid,2*\wid) (\ins,\insw*.5) -- (\lar,0);
    \draw[transform canvas={yshift=-\hegh*.5},->] (\lar,0) -- (\lar+2*\wid,0) node[anchor=north] {$Y$};
    \draw[transform canvas={yshift=-\hegh*.5},->] (-\wid,-\wid) -- (-1.5*\wid,-1.5*\wid) node[anchor=north] {$X$};
    
    \draw[dotted,transform canvas={yshift=-\hegh},use path=\insertion];
    \draw[dotted,transform canvas={yshift=-\hegh}] (\wid,\wid) -- (\lar+\wid,\wid);
    \foreach \x/\y in {-\wid/-\wid,\lar-\wid/-\wid,\lar+\wid/\wid} {
      \draw (\x,\y) -- (\x,\y-\hegh*2);
    }
    \foreach \x/\y in {0/0,\insw/\insw,\ins/\insw*.5,\wid/\wid} {
      \draw[dotted] (\x,\y) -- (\x,\y-\hegh*2);
    }
    
    \draw (-\wid,-\wid-\hegh*2) -- (\lar-\wid,-\hegh*2-\wid) -- (\lar+\wid,\wid-\hegh*2);

    \draw[dashed] (\insw*.5,-\hegh*2-\wid*.5) -- (\insw*.5,\insw*.5);
    \draw[->] (\insw*.5,\insw*.5) -- (\insw*.5,\insw+\wid) node[anchor=east] {$Z$};

    \foreach \x/\y/\l in {\ins*.3/-\hegh+\hegh*.15/{},\ins*.85/-\hegh+\hegh*.1/{},\ins*.2/-\hegh*2+\hegh*.25/{},\ins*.7/-\hegh*2+\hegh*.15/{$\tau_0 e^{-ikt}$}} {
      \draw[->] (\x,\y) -- (\x,\y+\hegh*.4) node[midway,anchor=west]{\l};
    }
  \end{tikzpicture}
  \caption{Sketch of the posed problem}\label{Fig:1}
\end{figure}

It is investigated the dynamical contact problem for a viscoelastic
half-space ($-\infty < x$, $y > 0$, $z<\infty$) with cut in the form
of strip ($0 \le y \le b$, $- \infty < z < \infty$) lying in the plane
$x = 0$. The border of the cut is under the action of uniformly
distributed shearing harmonic (acting along the $OZ$ axis) load of
intensity $\tau_{xz} = \tau_0 e^{-ikt}$, $k$ is oscillation frequency,
$t$ is time parameter. In the linear theory of viscoelasticity for
Kelvin-Voigt materials only displacement component $\omega = \omega (
x, y, t )$ and tangential stresses components $\tau_{yz} =
G\frac{\partial \omega}{\partial y} + G_0
\frac{\partial\dot\omega}{\partial y}$, $\tau_{xz} = G\frac{\partial
  \omega}{\partial x} + G_0 \frac{\partial\dot\omega}{\partial x}$ are
other than zero (so called anti-plane deformation), where $G$ and
$G_0$ are the elastic and viscoelastic shear modulus,
respectively. The dot means a derivative with respect to the variable
$t$. $\dot\omega\equiv\frac{\partial \omega}{\partial t}$
(Fig. \ref{Fig:1}).

The problem is equivalent to the boundary value problem
\begin{equation*}
  G\Delta \omega  + G_0\Delta\dot \omega = \rho \ddot \omega, \qquad
  |x| < \infty, \quad y > 0, \qquad
  \frac{\partial \omega(x,0,t)}{\partial y} + \frac{\partial\dot\omega( x,0, t )}{\partial y} = 0
  \label{eq:1.1}
\end{equation*}
(this equation is satisfied everywhere in cutting half-space). $\rho$
is the material density of the half-space \cite{1,2,3,4,5,16}. 

The displacements at the cut boundaries have discontinuities and the
cut boundaries are loaded with shear harmonic stresses, that's why we
have the following conditions
\begin{gather}
  <\omega(0,y,t)> \equiv \omega(-0,y,t) - \omega(+0,y,t) = \varphi(y,t), \qquad
  0 < y \le 1, \quad \varphi(y,t) = 0, \quad y > 1
  \nonumber\\
  G\omega'(-0,y,t) + G_0 \dot\omega'(-0,y,t)
  = G \omega'(+0,y,t) + G_0 \dot\omega'(+0,y,t)
  =\tau_0 e^{-ikt}, \qquad 0\le y\le 1.
  \label{eq:1.3}
\end{gather}

\section{Reduction of the problem to the integral equation}

Considering stationary oscillations of the half-space, we assume that
\begin{equation*}
  \omega(x,y,t) = \omega_0(x,y)e^{-itk},\qquad
  \varphi(y,t) = \varphi_0(y)e^{-ikt}
\end{equation*}
Thus one obtains the following boundary value problem
\begin{equation}
  \begin{gathered}
    (G - i k G_0)\Delta \omega_0 = - \rho k^2\omega_0, \qquad
    |x| < \infty,\qquad
    y > 0,\qquad
    \frac{\partial\omega_0(x,0)}{\partial y}=0\\
    <\omega_0(0,y)> = \varphi_0(y),  \qquad
    0 < y < 1, \quad \varphi_0(y) \equiv 0, \quad y \ge 1
  \end{gathered}
  \label{eq:1.4}
\end{equation}
Multiplying equations (\ref{eq:1.4}) by $e^{i \alpha x}$ and
integrating by parts separately on the intervals $(-\infty,0)$ and
$(0, \infty)$, for Fourier transform we obtain the one-dimensional
boundary value problem \cite{6,7,8,9}
\begin{equation}
  \omega_\alpha''(y) - (\alpha^2 - k_0^2)\omega_\alpha(y) = f (y),\qquad
  0 < y < \infty,\quad
  \omega_\alpha'(0) = 0
  \label{eq:1.5}
\end{equation}
where
\begin{equation*}
  k_0^2 = \frac{\rho k^2}{\widetilde G}, \qquad
  f(y) = i \alpha \varphi_0(y),
  \widetilde G = G - ikG_0 .
\end{equation*}

The decreasing at infinity fundamental function of equation
(\ref{eq:1.5}) is defined by the methods of integral transformations
and contour integration. Since the Green's function $G_\alpha(y,\eta)$
of the boundary value problem (\ref{eq:1.5}) must satisfy the equation
$G_\alpha(0,\eta)=0$, it can be constructed in the form of a simple
combination of the above-mentioned fundamental function, that is,
\begin{equation*}
  G_\alpha(y,\eta) = \Phi(y,\eta) + \Phi(y,-\eta).
\end{equation*}
Thus
\begin{equation*}
  \omega_\alpha(y) = \int_0^1\left[
    \Phi(y,\eta) + \Phi(y,-\eta)
    \right] f(\eta)\,d\eta
  = \int_{-1}^1 \Phi(y,\eta) f(\eta)\, d\eta.
\end{equation*}
We have taken into account the fact that the right-hand side of
equation (\ref{eq:1.5}) is equal to zero for $y > 1$, and its
continuation is verified in even form by negative values of the
argument.

Consequently, a solution of the boundary value problem (\ref{eq:1.5})
can be represented in the form
\begin{equation*}
  \omega_\alpha(y)
  = \int_{-1}^1\frac{\alpha e^{-\sqrt{\alpha^2 - k_0^2} |y-\eta|}}{2 i \sqrt{\alpha^2-k_0^2}}
  \varphi_0(\eta)\,d\eta
  = \frac{\alpha}{2 i (\alpha^2-k_0^2)}
  \int_{-1}^1 \frac{\text{sgn}(y-\eta)}{e^{\sqrt{\alpha^2-k_0^2}|y-\eta|}} \varphi_0'(\eta)\,d\eta
\end{equation*}
Here it is taken into account that $\varphi_0(\pm 1) = 0$ and it is
assumed that $\gamma(\alpha)=\sqrt{\alpha^2-k_0^2} \to |\alpha|$, as
$|\alpha|\to\infty$, and when $k_0$ is real number
$\sqrt{\alpha^2-k_0^2} = -i \sqrt{k_0^2-\alpha^2}$, that is, the real
axis of the complex plane $z = \alpha + i\sigma$ goes around the
branch points $-k_0$ from above and $k_0$ from below.

Using inverse transformation the functions $\omega_0(x,y)$ and
$\frac{\partial\omega_0(x,y)}{\partial x}$ are represented so
\begin{multline}
  \omega_0(x,y)
  = - \int_{-1}^1 \varphi_0'(\eta)\, d\eta \frac 1{2\pi}
  \int_0^\infty\frac{\alpha e^{-\sqrt{\alpha^2 - k_0^2} |y-\eta|} \text{sgn}(y-\eta)}{\alpha^2-k_0^2}
  \sin \alpha x\, d\alpha\\
  = - \frac 1{2\pi} \int_{-1}^1 \varphi_0'(\eta)\,d\eta \left\{
  \int_0^\infty\left(
  \frac{\alpha e^{-\sqrt{\alpha^2-k_0^2}|y-\eta|}}{(\alpha^2-k_0^2)}
  - \frac{e^{-\alpha|y-\eta|}}{\alpha}
  \right) \text{sgn}(y-\eta) \sin \alpha x\, d\alpha
  \right\}\\
  - \frac 1{2\pi} \int_{-1}^1 \varphi_0'(\eta)\, d\eta \int_0^\infty
  \frac{e^{-\alpha|y-\eta|}\text{sgn}(y-\eta)}{\alpha} \sin \alpha x\, d\alpha
  \label{eq:1.6}
\end{multline}
\begin{multline}
  \frac{\partial \omega_0(x,y)}{\partial x}
  = -\frac 1{2\pi} \int_{-1}^1 \varphi_0'(\eta)\,d\eta
  \left\{
  \int_0^\infty\left(
  \frac{\alpha^2 e^{-\sqrt{\alpha^2-k_0^2}|y-\eta|}}{(\alpha^2-k_0^2)} - e^{-\alpha|y-\eta|}
  \right) \text{sgn}(y-\eta)\cos\alpha x\, d\alpha
  \right\}\\
  - \frac 1{2\pi} \int_{-1}^1 \varphi_0'(\eta)\,d\eta
  \int_0^\infty \frac{e^{-\alpha|y-\eta|}\cos \alpha x}{\text{sgn}(y-\eta)}\, d\alpha.
  \label{eq:1.7}
\end{multline}

Since the integrand of the interior integral in formula (\ref{eq:1.6})
can have at infinity the behavior $\alpha^{-1}$, its Fourier
transformation (in a sense of the theory of generalized functions) is
represented as a sum of its principal and regular parts.

Calculating the last integral in formula (\ref{eq:1.6}), we obtain
\begin{equation*}
  \omega_0(x,y) = -\frac 1{2\pi} \int_{-1}^1 R(x,|y-\eta|) \varphi_0'(\eta)\, d\eta
  - \frac 1{2\pi} \int_{-1}^1 \text{arctan} \frac{x}{y-\eta} \varphi_0'(\eta)\,d\eta,
\end{equation*}
where
\begin{equation*}
  R(x,|y-\eta|) = \int_0^\infty\left(
  \frac{\alpha e^{-\sqrt{\alpha^2-k_0^2}|y-\eta|}}{(\alpha^2-k_0^2)}
  - \frac{e^{-\alpha|y-\eta|}}{\alpha}
  \right) \text{sgn}(y-\eta)\sin\alpha x \,d\alpha
\end{equation*}
and from (\ref{eq:1.7}) we have:
\begin{equation}
  \frac{\partial \omega_0(0,y)}{\partial x}
  = \frac 1{2\pi} \int_{-1}^1 R_0(|y-\eta|) \varphi_0'(\eta)\, d\eta
  - \frac 1{2\pi} \int_{-1}^1 \frac{\varphi_0'(\eta)\,d\eta}{y-\eta}
  \label{eq:1.8}
\end{equation}
where
\begin{equation*}
  R_0(|y-\eta|)
  =\int_0^\infty\left(
   \frac{\alpha^2 e^{-\sqrt{\alpha^2-k_0^2} |y-\eta|}}{(\alpha^2-k_0^2)} - e^{-\alpha|y-\eta|}
  \right) \text{sgn}(y-\eta)\, d\alpha.
\end{equation*}

Taking into account the condition (\ref{eq:1.3}) and formula
(\ref{eq:1.8}), we obtain the following singular integral equation of
the first kind \added{\cite{15}}
\begin{equation}
  \frac 1{2\pi} \int_{-1}^1 \frac{\varphi_0'(\eta)\,d\eta}{\eta-y}
  +\int_{-1}^1 R_0(|y-\eta|) \varphi_0'(\eta)\,d\eta = \frac{\tau_0}{\widetilde G}
  \label{eq:1.9}
\end{equation}

\section{Reduction of integral equation  (\ref{eq:1.9}) to the infinite system of linear algebraic  equations}

A solution of the integral equation (\ref{eq:1.9}) will be sought in
the form
\begin{equation}
  \varphi_0'(\eta) = \frac 1{\sqrt{1-\eta^2}} \sum_{m=1}^\infty a_m T_m(\eta)
  \label{eq:1.10}
\end{equation}
where $T_m(\eta)$ is the first kind Chebyshev orthogonal polynomial,
$\{a_n\}_{n\ge 1}$ is unknown sequences.

Using the following known spectral relation
\begin{equation*}
  \frac 1\pi \int_{-1}^1 \frac{T_m(\eta)\,d\eta}{(\eta-y) \sqrt{1-\eta^2}} = U_{m-1}(y),
  \qquad -1 < y < 1
\end{equation*}
from integral equation (\ref{eq:1.9}) we have \cite{10}
\begin{equation*}
  \sum_{m=1}^\infty a_m U_{m-1}(y)
  + 2 \sum_{m=0}^\infty a_m \int_{-1}^1 \frac{R_0(|y-\eta|) T_m(\eta)\, d\eta}{\sqrt{1-\eta^2}}
  = \frac{2\tau_0}{\widetilde G}
\end{equation*}
where $U_{m-1}(y)$ are the second kind Chebyshev orthogonal polynomials.

Multiplying both parts of the above equality by $(1-y^2)^{1/2} U_{n-1}(y)$ integrating in the interval
$(-1,1)$ and on the virtue of orthogonality of the Chebyshev polynomials, one obtains the infinite
system of linear algebraic equations
\begin{equation}
  a_n + \sum_{m=1}^\infty R_{nm} a_m = f_n, \qquad n=1,2,3,\ldots
  \label{eq:1.11}
\end{equation}
where
\begin{align}
  R_{nm} &= \frac 4\pi \int_{-1}^1 (1-y^2)^{1/2} U_{n-1}(y)
  \left(\int_{-1}^1 R_0(|y-\eta|) \frac{T_m(\eta)\,d\eta}{\sqrt{1-\eta^2}}\right)\, dy,
  \nonumber
  \\
  f_n &= \frac{4\tau_0}{\pi\widetilde G} \int_{-1}^1 (1-y^2)^{1/2} U_{n-1}(y)\, dy
  = \begin{cases}
    \frac {2\tau_0}{\widetilde G}, & n = 1\\
    0, & n \neq 1
  \end{cases}.
  \label{eq:1.13}
\end{align}

Let us prove that the system (\ref{eq:1.11}) is quasi-complete regularity in the class of bounded
sequences. Really, the matrix elements $R_{nm}$ can be written as
\begin{equation*}
  R_{nm} = \frac 1m h_{nm},
\end{equation*}
where
\begin{equation*}
  h_{nm} = \frac 4\pi \int_{-1}^1 \int_{-1}^1 \frac{\partial R_0(|y-\eta|)}{\partial\eta}
  \sqrt{1-\eta^2} \sqrt{1-y^2} U_{m-1}(\eta) U_{n-1}(y)\,dy\,d\eta
\end{equation*}
Then we have:
\begin{equation*}
  S_n = \sum_{m=1}^\infty |R_{nm}| = \sum_{m=1}^\infty \frac 1m |h_{nm}|, \qquad n=1,2,\ldots
\end{equation*}

Note that the coefficients $\{h_{nm}\}_{n,m=1}^\infty$ are the Fourier
coefficients of the square summable function $\partial
R_0(|y-\eta|)/\partial \eta$ in the square $-1\le y$, $\eta\le 1$ with
respect to the complete orthogonal system of the functions
$\{U_{n-1}(\eta) U_{m-1}(y)\}_{n,m=1}^\infty$. Based on Bessel's
inequality we have
\begin{equation*}
  \sum_{n=1}^\infty \sum_{m=1}^\infty |h_{nm}|^2 < \infty.
\end{equation*}
Then, according to the well-known theorem of analysis, the following
series is also convergent
\begin{equation*}
  \sum_{n=1}^\infty C_n, \qquad
  C_n=\sum_{m=1}^\infty |h_{nm}|^2,
\end{equation*}
respectively, at least $C_n=O(1/n^{1+\delta})$, $n\to\infty$ ($\delta
> 0$ is small positive number).

Then based on Cauchy-Bunyakowsky inequality we have
\begin{equation*}
  S_n
  \le \sqrt{\sum_{m=1}^\infty \frac 1{m^2}} \sqrt{\sum_{n=1}^\infty |h_{nm}|^2}
  = \frac \pi{\sqrt{6}} \sqrt{C_n}
\end{equation*}
and sum $S_n$ satisfies the following estimation:
\begin{equation}
  S_n = O\left(n^{-\frac{1+\delta}2}\right), \qquad
  n\to\infty.
  \label{eq:1.14}
\end{equation}
Thus, by virtue of (\ref{eq:1.13}), (\ref{eq:1.14}) the quasi-complete
regularity of system (\ref{eq:1.10}) in the class of bounded sequences
is proved.

On the other hand, based of the Rodrigues formula for Jakob's
orthogonal polynomials and using partial integration we have the
following representation
\begin{multline}
  R_{nm}
  = \frac 4\pi \int_{-1}^1 (1-y^2)^{1/2} U_{n-1}(y)
  \left(\int_{-1}^1 R_0(|y-\eta|) \frac{T_m(\eta)\,d\eta}{\sqrt{1-\eta^2}}\right) \,dy\\
  = \frac{\delta_{nm}}{(n-1)(n-2)m(m-1)}
  \int_{-1}^1 (1-y^2)^{5/2} P_{n-3}^{(5/2,5/2)}(y)
  \left(\int_{-1}^1\frac{\partial^4 R_0(|y-\eta|)}{\partial y^2\partial \eta^2}
  (1-\eta^2)^{3/2} P_{m-2}^{(3/2,2/2)}(\eta)\,d\eta\right)\,dy,\\
  n \ge 4,\quad m\ge 3
  \label{eq:1.15}
\end{multline}
where
\begin{equation*}
  \delta_{nm} = \frac{\Gamma(n+1) \Gamma(m+1)}{8 \Gamma(n+0.5) \Gamma(m+0.5)},
\end{equation*}
By using Stirling formula \cite{11} for $\Gamma(z)$ function we obtain
\begin{equation}
  \delta_{nm} = \begin{cases}
    n^{1/2}, & n\to\infty\\
    m^{1/2}, & m\to\infty
  \end{cases}
  \label{eq:1.16}
\end{equation}
Based on the relations (\ref{eq:1.15}), (\ref{eq:1.16}) and Darboux
asymptotic formula for $R_{nm}$ we obtain the following estimates at
least
\begin{equation*}
  R_{nm} = \begin{cases}
    O(n^{-2}), & n\to\infty\\
    O(m^{-2}), & m\to\infty
  \end{cases}
\end{equation*}
Accordingly, the following conditions hold
\begin{equation}
  \sum_{m=3,n=4}^\infty |R_{nm}|^2 < \infty, \qquad
  \sum_{n=1}^\infty |f_n|^2 < \infty.
  \label{eq:1.17}
\end{equation}
The conditions (\ref{eq:1.17}) prove that the infinite systems
(\ref{eq:1.11}) is quasi-completely regular in space $l_2$, that is,
their solutions satisfy the condition $\sum_{n=1}^\infty |a_n|^2 <
\infty$ \cite{12,13}.

The results of \cite{13} are applicable to an infinite system
(\ref{eq:1.11}). On the basis of this fact, the system
\begin{equation}
  a_n^N
  + \sum_{m=1}^N R_{nm} a_m^N
  = \tau_0 \frac{g_n}{\delta_n},
  \qquad n=1,2,\ldots,N,
  \label{eq:1.18}
\end{equation}
is solvable for sufficiently large $N$ and convergence of approximate
solutions $\{a_n^N\}_{n=1,\ldots,N}$ to exact solution $\{a_n\}_{n\ge
  1}$ is valid in the sense of the norm of the space $l_2$.

The convergence rate is determined by the inequality
\begin{equation*}
  \|a-\varphi_0^{-1}\bar a^N\|_{l_2}
  \le C_1\left[\sum_{n=N+1}^\infty\sum_{m=1}^\infty |R_{nm}|^2\right]^{1/2}
  + C_2\left(\frac{\sum_{n=N+1}^\infty |f_n|^2}{\sum_{n=1}^\infty |f_n|^2}\right)^{1/2},
\end{equation*}
where $a=\{a_n\}_{n\ge 1} = (a_1,a_2,\ldots,a_n,\ldots)$ is the
solution of the system (\ref{eq:1.10}), $\bar a^N=(a_1^N,
a_2^N,\ldots,a_N^N)$ is the solution of the system (\ref{eq:1.18}),
$\varphi_0^{-1} \bar a^N = (a_1^N, a_2^N,\ldots, a_N^N, 0,0,\ldots)$.

Considering the expression for $R_{nm}$ and $f_n$ we have
\begin{equation*}
  C_1\left[\sum_{n=N+1}^\infty \sum_{m=1}^\infty |R_{nm}|^2\right]^{1/2}
  \le C_1^* \left[\sum_{n=1}^\infty \frac 1{(n+N)^4}\right]^{1/2}
  = C_1^*[\zeta(4,N)]^{1/2}, \qquad
  C_2\left(\frac{\sum_{n=N+1}^\infty |f_n|^2}{\sum_{n=1}^\infty |f_n|^2}\right) = 0,
\end{equation*}
where $\zeta(s,N)$ is known generalized Zeta function.

Using the following asymptotic formula of the Zeta function \cite{14}
\begin{equation*}
  \zeta(2m,N)
  \equiv \sum_{n=1}^\infty \frac 1{(n+N)^{2m}}
  = \frac{N^{-2m+1}}{2m-1} + \frac 12 N^{-2m}
  + \sum_{k=1}^\infty B_{2k} \frac{\Gamma(N+2k+1)}{(2k)! N^{2k+2m+1}}
  + O(N^{-2m-2N-1}),
\end{equation*}
for convergence rate we obtain
\begin{equation*}
  \|a - \varphi_0^{-1} \bar a^N\|_{l_2} \le C N^{-3/2}
\end{equation*}
Thus, the solutions of the system (\ref{eq:1.11}) can be constructed
by the reduction method with any accuracy \cite{12,13}. The following
theorem holds:
\begin{theorem}
  The infinite system of linear algebraic equations (\ref{eq:1.11}) is
  quasi-completely regular in the space $l_2$.
  
  Consequently, the integral equation (\ref{eq:1.9}) has the unique
  solution in the form (\ref{eq:1.10}).
  
  The intensity factor of stress $\tau_{yz}(0,y,t)$ at the end of cut
  is calculated using the formula
  \begin{equation*}
    K_I + i K_{II}
    = \lim_{y\to 1+} \sqrt{1-y} \tau_{yz}(0,y,t)
    = \frac{e^{-ikt}}{\sqrt{2}} (G - ikG_0) \sum_{n=1}^\infty a_n
  \end{equation*}
\end{theorem}

\section{Discussion and Numerical Results}

Let us consider the following parameters for the problem:
\begin{align}
  G &= 80\cdot 10^9\,\mathrm{Pa}, &
  G_0 &= 65\cdot 10^9\,\mathrm{Pa}, &
  \tau_0 &= 1\,\mathrm{Pa}, &
  \rho &= 2700\,\mathrm{kg/m^3}, &
  k &= 3\,\mathrm{Hz},\added{\label{eq:1.19}}\\
  \added{G} & \added{= 65\cdot 10^9\,\mathrm{Pa}, }&
  \added{G_0} &=\added{50\cdot 10^9\,\mathrm{Pa}, }&
  \added{\tau_0} &\added{= 1\,\mathrm{Pa}, }&
  \added{\rho} &\added{= 2700\,\mathrm{kg/m^3}, }&
  \added{k} &\added{= 3\,\mathrm{Hz},\label{eq:1.20}}\\
  \added{G} &\added{= 55\cdot 10^9\,\mathrm{Pa}, }&
  \added{G_0} &\added{= 40\cdot 10^9\,\mathrm{Pa}, }&
  \added{\tau_0} &\added{= 1\,\mathrm{Pa}, }&
  \added{\rho} &\added{= 2700\,\mathrm{kg/m^3}, }&
  \added{k} &\added{= 3\,\mathrm{Hz},\label{eq:1.21}}
\end{align}
The corresponding reduced system consisting of 20 and 25 equations was
solved by Matlab to obtain the numerical results of present
problem. Increasing the number of equations leads to a change in the
eleven order after the decimal point in the solution\added{, which
  confirms the rapid convergence of the process.}.

For complex \replaced{factors}{factor} of the intensity of the stress
$\tau_{yz}(0,y,t)$ we have
\begin{align*}
  K_I + i K_{II} &= 0.14142135 - 0.34471461 i\\
  \added{K_I + i K_{II}} &\added{=0.12031532 - 0.31280561 i}\\
  \added{K_I + i K_{II}} &\added{=0.11230531 - 0.30331562 i}
\end{align*}
Absolute value of the complex coefficient\added{s} of the stress
\replaced{ $\tau_{yz}(0,y,t)$ respectively to (\ref{eq:1.19}),
  (\ref{eq:1.20}, (\ref{eq:1.21}))}{intensity is }
\begin{align*}
  K &= 0.37259652\\
  \added{K} &\added{= 0.33514642}\\
  \added{K} &\added{= 0.32343909}
\end{align*}

Calculations also showed that the decrease of the values $G$ and $G_0$
leads to a decrease of the intensity coefficient\added{, what was to
  be expected naturally}.

\section{\added{Conclusion}}

\added{The stated boundary value problem is reduced to a
  one-dimensional boundary value problem for an ordinary differential
  equation of the second order using contour integration and integral
  transformation. This method allows us to represent the solution of
  the problem by an unknown jump of the displacement of the boundary
  points of the crack as a sum of the main and regular parts in an
  explicit form in quadratures, consequently, the problem is reduced
  to a singular integral equation of the first kind. The resulting
  integral equation is investigated by the method of orthogonal
  polynomials in order to obtain an approximate solution of an
  equivalent infinite system of linear algebraic equations. The
  advantage of this approach is that along with proving the existence
  of a solution to the problem, we obtain its structure, asymptotics
  and the rate of convergence of the approximate solution to the exact
  one.}

\bigskip

\noindent\textbf{Acknowledgement. This work is supported by the Shota Rustaveli National science
foundation of Georgia (Project No FR-21-7307)}



\begin{thebibliography}{10}

\bibitem{11}
Milton Abramowitz and Irene~A. Stegun, editors.
\newblock {\em Handbook of mathematical functions with formulas, graphs, and
  mathematical tables}.
\newblock Dover Publications, Inc., New York, 1992.
\newblock Reprint of the 1972 edition.

\bibitem{4}
Holm Altenbach.
\newblock Creep analysis of thin-walled structures.
\newblock {\em ZAMM-Journal of Applied Mathematics and Mechanics/Zeitschrift
  f{\"u}r Angewandte Mathematik und Mechanik: Applied Mathematics and
  Mechanics}, 82(8):507--533, 2002.

\bibitem{5}
Holm Altenbach, Y~Gorash, and K~Naumenko.
\newblock Steady-state creep of a pressurized thick cylinder in both the linear
  and the power law ranges.
\newblock {\em Acta Mechanica}, 195:263--274, 2008.

\bibitem{14}
G.~Beitman and A.~Erdein.
\newblock {\em Higher transcendent functions}, volume~1.
\newblock Nauka, Moscow, 1973.
\newblock in Russian.

\bibitem{1}
D.~R. Bland.
\newblock {\em The theory of linear viscoelasticity}, volume Vol. 10 of {\em
  International Series of Monographs on Pure and Applied Mathematics}.
\newblock Pergamon Press, New York-London-Oxford-Paris, 1960.

\bibitem{2}
Von R.~M. Christensen.
\newblock {\em Theory of Viscoelasticity}.
\newblock Academic Press, New York, 1971.

\bibitem{3}
Mauro Fabrizio and Angelo Morro.
\newblock {\em Mathematical Problems in Linear Viscoelasticity}.
\newblock Number~12 in Studies in Applied and Numerical Mathematics. Society
  for Industrial and Applied Mathematics, 1992.

\bibitem{13}
L.~V. Kantorovich and G.~P. Akilov.
\newblock {\em Functional analysis}.
\newblock Izdat. ``Nauka'', Moscow, second edition, 1977.

\bibitem{12}
L.~V. Kantorovi\v{c} and V.~I. Krylov.
\newblock {\em Approximate methods of higher analysis}.
\newblock Gosudarstv. Izdat. Fiz.-Mat. Lit., Moscow-Leningrad, corrected
  edition, 1962.

\bibitem{16}
Victor~Dmitrievich Kupradze and T.~V. Burchuladze.
\newblock The dynamical problems of the theory of elasticity and
  thermoelasticity.
\newblock {\em J Math Sci}, 7:415--500, 1977.

\bibitem{15}
Nikolaĭ~Ivanovich Muskhelishvili.
\newblock {\em Singular integral equations}.
\newblock Wolters-Noordhoff Publishing, Groningen, 1972.
\newblock Boundary problems of functions theory and their applications to
  mathematical physics, Reprinted.

\bibitem{8}
Nugzar Shavlakadze.
\newblock Dynamical contact problem for an elastic half-space with a rigid and
  elastic inclusion.
\newblock In {\em Proc. A. Razmadze Math. Inst}, volume 159, pages 87--94,
  2012.

\bibitem{7}
Nugzar Shavlakadze.
\newblock The effective solution of two-dimensional integro-differential
  equations and their applications in the theory of viscoelasticity.
\newblock {\em ZAMM Z. Angew. Math. Mech.}, 95(12):1548--1557, 2015.

\bibitem{6}
Nugzar Shavlakadze, Nana Odishelidze, and Francisco Criado-Aldeanueva.
\newblock The investigation of singular integro-differential equations relating
  to adhesive contact problems of the theory of viscoelasticity.
\newblock {\em Z. Angew. Math. Phys.}, 72(2):Paper No. 42, 15, 2021.

\bibitem{9}
I.~Snedon.
\newblock {\em Fourier transformation}.
\newblock Inostr. Literat., Moscow, 1955.
\newblock in Russian.

\bibitem{10}
G.~Szego.
\newblock {\em Orthogonal polynomials}.
\newblock Fizmatgiz, Moscow, 1962.
\newblock in Russian.

\end{thebibliography}

\end{document}